# Covert Channels in SIP for VoIP signalling

Wojciech Mazurczyk, Krzysztof Szczypiorski

Warsaw University of Technology, Faculty of Electronics and Information
Technology, Institute of Telecommunications, 15/19 Nowowiejska Str.
00-665 Warsaw, Poland
`{W.Mazurczyk, K.Szczypiorski}@tele.pw.edu.pl`

**Abstract.** In this paper, we evaluate available steganographic techniques for SIP (Session Initiation Protocol) that can be used for creating covert channels during signaling phase of VoIP (Voice over IP) call. Apart from characterizing existing steganographic methods we provide new insights by introducing new techniques. We also estimate amount of data that can be transferred in signalling messages for typical IP telephony call.

## 1 Introduction

Steganography is a process of hiding secret data inside other, normally transmitted data. Usually, it means hiding of a secret message within an ordinary message and its extraction at the destination point. In an ideal situation, anyone scanning this information will fail to know whether it contains covert data or not. A covert channel [9] is one of the most popular steganographic techniques that can be applied in the networks. The covert channel offers an opportunity to "manipulate certain properties of the communications medium in an unexpected, unconventional, or unforeseen way, in order to transmit information through the medium without detection by anyone other than the entities operating the covert channel" [17].

Nowadays, VoIP is one of the most popular services in IP networks. It stormed into the telecom market and changed it entirely. As it is used worldwide more willingly, the traffic volume that it generates is still increasing. That is why VoIP traffic may be used to enable hidden communication throughout IP networks. Applications of the VoIP covert channels differ as they can pose a threat to the network communication or can be used to improve the functioning of VoIP (e.g. security like in [11] or quality of service like in [10]). The first application of the covert channel is more dangerous as it can lead to the confidential information leakage. It is hard to assess what bandwidth of a covert channel poses a serious threat, it depends on the security policy that is implemented in the network. For example: The US Department of Defense specifies in [16] that any covert channel with bandwidth higher than 100 bps must be considered insecure for average security requirements. Moreover for high security requirements it should not exceed 1 bps.

In this paper we present available covert channels that may be utilized for hidden communication for SIP protocol used as a signalling protocol for VoIP service. Moreover, we introduce new steganographic methods that, to our best knowledge, were not described earlier. For each of these methods we estimate potential bandwidth to later evaluate how much information may be transferred in a typical IP telephony call.

The paper is organized as follows. In Section 2 we circumscribe the types of VoIP traffic and a general communication flow for IP telephony call. In Section 3, we describe available steganographic methods that may be used to create covert channels for signalling messages. Then, in Section 4, we estimate a total amount of data that may be transferred with use of the SIP protocol. Finally, Section 5 concludes our work.

## 2 VoIP Communication Flow

VoIP is a real-time service that enables voice conversations through IP networks. Protocols that are used for creating IP telephony may be divided into four following groups:

a. *Signalling protocols* which allow to create, modify, and terminate connections between the calling parties. Nowadays the most popular are SIP [14], H.323 [6], and H.248/Megaco [3],
b. *Transport protocols,* from which the most important one is RTP [15], which provides end-to-end network transport functions suitable for applications transmitting real-time audio. RTP is used in conjunction with UDP (or rarely TCP) for transport of digital voice stream,
c. *Speech codecs* e.g. G.711, G.729, G.723.1 that allow to compress/decompress digitalized human voice and prepare it for transmitting in IP networks,
d. Other *supplementary protocols* like RTCP [15], SDP [5], or RSVP etc. that complete VoIP functionality. For purposes of this paper we explain the role of SDP protocol, which is used with SIP messages to describe multimedia sessions and to negotiate their parameters.

IP telephony connection may be divided into two phases: a *signalling phase* and a *conversation phase*. In both of these phases certain types of traffic are exchanged between calling parties. In this paper we consider VoIP service based on the SIP signaling protocol (with SDP) and RTP (with RTCP as control protocol) for audio stream transport. It means that during the signalling phase of the call certain SIP messages are exchanged between SIP endpoints (called: SIP User Agents). SIP messages usually traverse through SIP network servers: proxies or redirects that help end-users to locate and reach each other. After this phase, a conversation phase begins, where audio (RTP) streams flow bi-directly between a caller and a callee. VoIP traffic flow described above and distinguished phases of the call are presented in Fig. 1. For more clarity, we omitted the SIP network servers in this diagram (as they interpret the signalling messages and can modify only a few fields of SIP message which we will not use for steganographic purposes). Also potential security mechanisms in traffic exchanges were ignored.

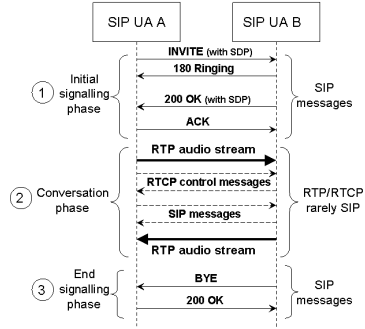

**Fig. 1.** VoIP call setup based on SIP/SDP/RTP/RTCP protocols (based on [7])

## 3 VoIP Signalling Covert Channels Overview and New Insights

In this section we will provide an overview of existing and new steganographic techniques used for creation of covert channels for VoIP that may be used during signalling phase of the call. To calculate potential amount of information that may be exchanged between calling parties we define *total amount of covert data ($B_T$)* that refers to information transferred (in bits) in SIP signalling messages (in one direction) with the use of all described steganographic methods. It can be expressed as:

$$B_T = \sum_{j=1}^{k} B_j \qquad (1)$$

where: $B_j$ describes amount of covert data transferred with use of the covert channel created by each steganographic method used during VoIP signalling and $k$ is a number of steganographic techniques used for VoIP signalling.

Traffic generated during the signalling phase of the call is provided from SIP signalling messages that are exchanged between both endpoints. That is why, we can point out the following steganographic methods to create covert channels:
- *TCP/UDP/IP* steganography in transport and network layers of TCP/IP stack,
- *SIP/SDP* protocols steganography in application layer of TCP/IP stack.

### 3.1 IP/TCP/UDP Protocols Steganography
TCP/UDP/IP protocols steganography utilizes the fact that only few fields of headers in the packet are changed during communication process ([12], [1], [13]). Covert data is usually inserted into redundant fields (provided, but often unneeded) for abovementioned protocols and then transferred to the receiving side. In TCP/IP stack, there are a number of methods available, whereby covert channels may be established and data can be exchanged between communication parties secretly. An analysis of the headers of typical TCP/IP protocols e.g. IP, UDP, TCP, but also e.g. HTTP (Hypertext Transfer Protocol) or ICMP (Internet Control Message Protocol) results in fields that are either unused or optional [1], [18]. This reveals many possibilities where data may be stored and transmitted. As described in [12] the IP header possesses fields that are available to be used as covert channels. The total capacity of those fields is rather high (as for the steganographic technique) and may exceed 32 bits per packet and there are also fields of TCP and UDP protocols that can be also used for this purpose. Notice that this steganographic method plays an important role for VoIP communication because protocols mentioned above are present in every packet (regardless, if it is a signalling message, audio packet, or control message).

**3.2 SIP/SDP Protocols Steganography**

To our best knowledge little research effort has been made to use SIP messages as a covert channel. For example in [2] authors have shown how the bouncing mechanism is used for SIP messages to secretly transfer data. The interest of research in SIP/SDP protocols steganography is rather low because the signalling phase is rather short and only few messages are exchanged during this phase. In spite of this observation we want to perform an analysis of covert channels that may be utilized for SIP signalling protocol to show how much information may be transferred in VoIP signalling messages – as mentioned in Section 1 transferring even 1 bps may be considered as a threat. When call setup begins, certain SIP signalling messages are exchanged between calling parties as depicted in Fig. 1 (marked as 1). Exemplary SIP message (with SDP session description) looks as presented in Fig. 2.

```
(1)     INVITE sip:bob@biloxi.example.com SIP/2.0
(2)     Via: SIP/2.0/TCP client.atlanta.example.com:5060;branch=z9hG4bK74bf9
(3)     Max-Forwards: 70
(4)     From: Alice <sip:alice@atlanta.example.com>;tag=9fxced76sl
(5)     To: Bob <sip:bob@biloxi.example.com>
(6)     Call-ID: 3848276298220188511@atlanta.example.com
(7)     CSeq: 12345 INVITE
(8)     Contact: AliceM <sip:alice@client.atlanta.example.com;transport=tcp>
(9)     Content-Type: application/sdp
(10)    Content-Length: 151

(11)        v=0
(12)        o=alice 2890844526 2890844526 IN IP4 client.atlanta.example.com
(13)        s=-
(14)        c=IN IP4 192.0.2.101
(15)        t=0 0
(16)        k=clear:9123123kjnhdasdoq12e31021n2e4
(17)        m=audio 49172 RTP/AVP 0
(18)        a=rtpmap:0 PCMU/8000
```
**Fig. 2.** Exemplary SIP INVITE signalling message with SDP session description
(bolded are fields and tokens that can be used for covert transmission)

First part of the message in Fig. 2 (signalling message header – marked with grey filling) is a SIP INVITE message (which initiates a call), the second part is an SDP session description (body of the message – marked with white filling).

**3.2.1 SIP Parameters, Tokens and Fields Steganography**

In SIP signalling messages there are certain tokens, like *tag* (in *From* field line 4, that forms SIP dialog identifier) or *branch* (in *Via* field line 2 that forms transaction identifier). They consist of random strings generated by user's endpoint when the connection is initiated. Also the fields: *Call-ID* (line 6, which uniquely identifies a call) and first part of *CSeq* field (line 7, initial sequence number that serves as a way to identify and order transactions) must be generated randomly. All abovementioned fields and tokens can be straightforwardly utilized as a low-bandwidth, one direction covert channel. However, for tag token [14] it stands that "when a tag is generated (…) it must be globally unique and cryptographically random with at least 32 bits of randomness…" – that means that the inserted secret value must be chosen appropriately. For value of a branch token the situation is similar, it must begin with the characters "z9hG4bK" (called magic cookie) to ensure that previous, older SIP version's implementation would not pick such a value. The rest of branch content is implementation-defined. Next, *Call-ID* is generated by the combination of a random string and the endpoint's host name or IP address (*random_string@host_name*). Moreover *CSeq* field consists of a sequence number and a method name; sequence number value, which is chosen arbitrarily, may be used for covert transmission. The only requirement for this number is that it must be expressible as a 32-bit unsigned integer and must be less than $2^{31}$. For all of the mentioned tokens and fields there are no rules inside a SIP standard (besides for *CSeq*) that specify their length, so we can increase the bandwidth of the covert channel by choosing appropriate length of those values. There is also a field *Max-Forwards* (line 3), that is used for loop detection. It may be also used as a covert channel, if the value applies to certain rules: SIP standard defines only that the initial value of *Max-Forwards* should be 70, however other values are also allowed. Eventually, we can also utilize strings in SIP messages e.g. in *Contact* field (line 8) – a string AliceM. Such string values have no direct impact on the communication itself. Fields that can be exploited in the same way as *Contact* include (more rarely, not mandatory) fields like: *Subject, Call-Info, Organization, Reply-To, Timestamp, User-Agent*, and other.

### 3.2.2 SIP Security Mechanisms Steganography

For SIP/SDP protocols steganography we can also utilize security mechanisms that are executed to provide security services like authentication and confidentiality for signalling messages. Especially end-to-end mechanisms are important for our purposes as they allow to transfer data directly between end users. In this article we will present how to use end-to-end SIP security mechanism S/MIME (Secure MIME) [4] to create covert channel. Fig. 3 presents how the SDP content, embedded into the SIP INVITE message, may be encrypted and signed using S/MIME. The secured parts of the message are divided from themselves using boundary value (*992d915fef419824* value in Fig. 3). It is the first value that can be utilized as a covert channel as its length and value is chosen randomly. Next, the first part between the boundary values is the *application/pkcs7-mime* binary *envelopedData* structure that encapsulates encrypted SDP session description. The second part between the boundary values is a signature of the payload (*application/pkcs7-signature*).

The second possibility for hidden communication is to use the signature bits inside the boundary values (*application/pkcs7-signature*) to transfer covert data. Therefore, we resign from signature verification (it is the cost of using this method), but instead, we gain an opportunity to send additional covert data. The amount of data that can be transferred covertly depends on what hash function is used and must be matched properly.

```
(1)     INVITE sip: bob@biloxi.example.com SIP/2.0
(2)     Via: SIP/2.0/UDP 160.85.170.139:5060;branch=z9hG4bK4129d28b8904
(3)     To: Bob <sip: bob@biloxi.example.com>
(4)     From: Alice <sip: alice@atlanta.example.com>;tag=daa21162
(5)     Call-ID: 392c3f2b568e92a8eb37d448886edd1a@160.85.170.139
(6)     CSeq: 1 INVITE
(7)     Max-Forwards: 70
(8)     Contact: <sip:alice@client.atlanta.example.com:5060>
(9)     Content-Type: multipart/signed;boundary=992d915fef419824;
(11)    micalg=sha1;protocol=application/pkcs7-signature
(12)    Content-Length: 3088
(13)    --992d915fef419824
(14)    Content-Type: application/pkcs7-mime;
(15)    smime-type=envelopeddata; name=smime.p7m
(16)    Content-Disposition: attachment;handling=required;filename=smime.p7m
(17)    Content-Transfer-Encoding: binary
(18)    <envelopedData object encapsulating encrypted SDP attachment not shown>
(19)    --992d915fef419824
(20)    Content-Type: application/pkcs7-signature; name=smime.p7s
(21)    Content-Transfer-Encoding: base64
(22)    Content-Disposition: attachment; filename=smime.p7s;
(23)    handling=required
(24)
(25)    ghyHhHUujhJhjH77n8HHGTrfvbnj756tbB9HG4VQpfyF467GhIGfHfYT6
(26)      QpfyF467GhIGfHfYT6jH77n8HHGghyHhHUujhJh756tbB9HGTrfvbnj
(27)    n8HHGTrfvhJhjH776tbB9HG4VQbnj7567GhIGfHfYT6ghyHhHUujpfyF4
(28)          7GhIGfHfYT64VQbnj756
(29)
(30)    --992d915fef419824--
```

**Fig. 3.** Example of SIP INVITE signalling message secured with S/MIME mechanism

### 3.2.3 SDP Protocol Steganography

For SDP protocol available covert channels are similar to those presented for SIP. In Fig. 1 SDP session description is enclosed in two SIP messages (INVITE from SIP UA A to SIP UA B and in 200 OK response in the reverse direction). It is possible to use session description fields in SDP protocol, some of them do not carry important information and other are ignored (but must be present in SIP/SDP message in order to be compliant with SDP). This includes bolded fields in Fig. 2 (second part with white filling): *v* (version – field ignored by SIP), *o* (owner/creator) – there is a randomly generated session identifier (*2890844526*), and the name of the owner/creator, *s* (session name – field ignored by SIP), *t* (time session is active – field ignored by SIP) and *k* (potential encryption key if the secure communication is used).

To summarize: for SIP/SDP protocols steganography creation of covert channels is possible because in specifications of these protocols there are no strict rules how to generate tokens and parameters and what is their desired length.

### 3.2.4 Other SIP/SDP Protocol Steganography Possibilities

For both protocols other steganographic methods may be utilized. For example, like in [8] we can use nonprintable characters (like spaces [SP] or tabs [HT]) or their sequences after the SIP header fields. Described situation is presented in Fig. 4.

The next method from [8] exploits the fact that the order of headers in the SIP/SDP message depends on implementation, thus reordering of headers is possible as a mean to covertly send data. If we consider exemplary signalling message form Fig. 4, if field *Call-ID* is after *CSeq* it can denote that binary "1" was sent, while if the order is reversed the value is "0". The last method exploits case modification (upper and lower cases), because names of the field are case-insensitive (so e.g. **FROM** header means "1" while *to* header "0"), but this technique is rather easy to uncover.

While call lasts, some signaling messages may also be exchanged to influence certain parameters of the session (e.g. codec). Bandwidth and steganographic techniques for SIP/SDP remain the same as described in the signalling phase. Moreover, during the conversation phase, we can also utilize SIP message like OPTIONS, which is used for sharing capabilities of the endpoints, e.g. to be able to support additional services. Such messages may be intentionally invoked (to some extent) to increase the covert channel bandwidth for these steganographic techniques. It is also worth noting that the SIP signalling messages are exchanged after the conversation phase is finished (marked on Fig. 1 with 3).

```
(1)    INVITE[SP]sip:bob@biloxi.example.com[SP]SIP/2.0[SP][SP][HT][SP][HT]
(2)    From: Alice <sip:alice@atlanta.example.com>;tag=9fxced76sl[HT][SP][HT]
(3)    To: Bob <sip:bob@biloxi.example.com>[HT][SP][HT][HT][SP][HT][SP][SP]
(4)    Call-ID: 3848276298220188511@atlanta.example.com[SP][HT][SP][SP]
(5)    CSeq: 12345 INVITE
```

**Fig. 4.** Example of usage of nonprintable characters as a covert channel for SIP

## 4 Evaluation of Total Covert Data Transferred in VoIP Signalling

Let us consider a scenario from Fig. 1 and based on that we will try to estimate how much information one may hide in signalling messages during the VoIP call. From Fig. 1 we can conclude that about 5 signalling messages may be sent in one direction between end users (two during initial signalling phase, two during the conversation e.g. OPTIONS message and one to end the call). Moreover, let us assume that two of these messages will carry also SDP body and that:

- IP/TCP/UDP protocols steganography provides covert transmission at the rate of 16 bits/message,
- SIP parameters, tokens and fields steganography gives about 60 characters for the first SIP message that is total of 480 bits (usage of initial values),
- SIP security mechanisms steganography which provides 160 bits per SIP message,
- SDP protocol steganography that gives 60 characters for each SDP body (we assumed two SDP bodies) that result in total of 960 bits,
- Other SIP/SDP protocol steganography possibilities we assumed about 8 bits/message.

For the considered scenario from Fig. 1 and equation 1 we can easily calculate that $B_T = 2.36$ kbits. Therefore, we see that even for only five SIP messages exchanged during VoIP call we can covertly transfer, in one direction, more than two thousand bits. That is why for high security requirements networks we may consider SIP steganography as a potential threat to information security.

## 5 Conclusions

In this paper we have described existing and introduced new steganographic methods for SIP/SDP protocols. All new solutions are based on network steganography as they utilized free or unused fields in abovementioned protocols. Total amount of information that may be transferred with use of proposed solutions is more than 2000 bits in one direction for each performed VoIP call. Although, this amount of information may be considered as low (as not many SIP/SDP messages are exchanged between end users), sometimes even this amount of data may be sufficient to cause serious information leakage.